\documentclass[showpacs,prl,floatfix,amsmath,amsfonts,superscriptaddress,twocolumn]{revtex4}
\usepackage{natbib,graphicx}
\usepackage[dvipdf]{hyperref}
\newcommand{\pd}[2]{\frac{\partial #1}{\partial #2}}
\newcommand{\pdd}[2]{\frac{\partial^2 #1}{\partial #2^2}}
\newcommand{\abs}[1]{\left\lvert#1\right\rvert}
\newcommand{\vecM}{\vec{M}}
\newcommand{\vm}{\vec{m}}
\newcommand{\MS}{M_s}

\newcommand{\ud}{\mathrm{d}}
\DeclareMathOperator{\re}{Re}
\DeclareMathOperator{\im}{Im}

\begin{document}
\title{Theory of Magnetodynamics Induced by Spin Torque in
  Perpendicularly Magnetized Thin Films}
\date{\today}
\author{M. A. \surname{Hoefer}}
\email{hoefer@colorado.edu}
\author{M. J. \surname{Ablowitz}}
\author{B. \surname{Ilan}}
\affiliation{Department of Applied Mathematics, University of
 Colorado, Boulder, Colorado 80309-0526, USA}
\author{M. R. \surname{Pufall}}
\thanks{Contribution of the U.S. Government.  Not subject to copyright.}
\author{T. J. \surname{Silva}}
\thanks{Contribution of the U.S. Government.  Not subject to copyright.}
\affiliation{National Institute of Standards and Technology, Boulder,
  Colorado 80305, USA}
\begin{abstract}
  A nonlinear model of spin wave excitation using a point contact in a
  thin ferromagnetic film is introduced.  Large-amplitude magnetic
  solitary waves are computed, which help explain recent spin-torque
  experiments.  Numerical simulations of the fully nonlinear model
  predict excitation frequencies in excess of 0.2 THz for contact
  diameters smaller than 6 nm.  Simulations also predict a saturation
  and red shift of the frequency at currents large enough to invert
  the magnetization under the point contact.  The theory is
  approximated by a cubic complex Ginzburg-Landau type equation.  The
  mode's nonlinear frequency shift is found by use of perturbation
  techniques, whose results agree with those of direct numerical
  simulations.
\end{abstract}
\maketitle

\begin{figure}[h]
  \centering
  {\includegraphics[scale=.39]{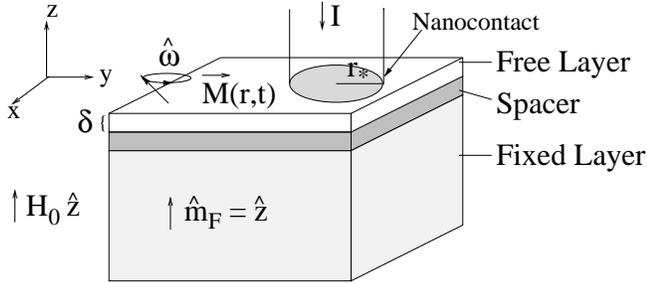}}
  \caption{Magnetic multilayer with magnetization $\vec{M}$ precessing
    in the free layer with frequency $\hat{\omega}$.  The
    magnetization of the lower ferromagnetic layer is fixed with
    orientation $\hat{m}_F$.  A nonmagnetic spacer is sandwiched
    between two ferromagnetic layers.  Typical parameters are: $r_* =
    20$ nm and $\delta = 5$ nm.}
  \label{fig:experiment}
\end{figure}

Generation of spin waves at microwave frequencies due to spin-momentum
transfer (SMT) in a thin magnetic multilayer is paving the way to a
new frontier in magnetodynamics.  Initial predictions, in the small
amplitude limit, made by Slonczewski \cite{Slon96} were recently
validated experimentally \cite{Tsoi98,Kiselev03,Silva04}.  In refs
\cite{Silva04,Tsoi98}, microwaves were generated by the application of
a dc current through a nanocontact into a trilayer of alternating
ferromagnetic and nonmagnetic materials (see fig
\ref{fig:experiment}).  Several models
\cite{Berger96,Stiles02,Bazaliy98} have been suggested to explain this
phenomenon.  More recent work has neglected spatial variation
\cite{Rezende05,Bertotti05,Slavin05}.  In this study, a theory of
\emph{nonlinear spatially nonuniform} magnetodynamics is developed and
analyzed.  Microwave emission from point contacts has potential
applications in communications and fast, high-density magnetic data
storage.

The system depicted in fig. \ref{fig:experiment} has been shown
experimentally to give rise to high-frequency microwaves inferred to
be caused by steady-state precession of magnetization in the
ferromagnetic free layer ($\text{Ni}_{80}\text{Fe}_{20}$)
\cite{Silva04}.  An external magnetic field $\vec{H}_0$ is oriented
perpendicular to the film plane.  When a dc current, $I$, is applied
through a point contact into the multilayer
$\text{Co}_{90}\text{Fe}_{10}$/Cu/$\text{Ni}_{80}\text{Fe}_{20}$, the
conduction electrons' spins induce a torque on the magnetization in
the free layer due to conservation of angular momentum
\cite{Slon96,Berger96}.  This SMT torque opposes the intrinsic damping
of the free layer and produces complex dynamics.  Slonczewski's linear
theory of spin wave excitation for point contacts on magnetic
multilayers predicts a threshold current for the excitation of steady
precession of magnetization \cite{Slon99}.  However, it does not
accurately predict the high frequencies seen in large amplitude
experiments nor the dependence of the frequency on current.

Understanding the SMT effect in point contacts with a perpendicular
geometry is of great interest as it is a microwave source that yields
the narrowest line widths and highest output powers
\cite{Silva04,Rippard05} when compared to other SMT geometries, such
as nanopillars \cite{Kiselev03} and in-plane fields \cite{Rippard04}.
In addition, this configuration is most amenable to analytical study
given the high symmetry of the problem.  Insights derived from this
simpler geometry may then be employed in solving the more advanced
problems of in-plane or oblique angled fields.

In this Letter, beginning with a nonlinear vectorial equation, we
derive a new complex cubic Ginzburg-Landau (CGL) equation for weakly
nonlinear excitations.  The dipole coupling and exchange
nonlinearities are shown to be responsible for the higher frequencies.
Steady-state precessing modes for the nonlinear vectorial model are
found via numerical integration.  The calculated frequencies lie in
the range of experimental observations \cite{Silva04}.  Using this
model, much higher frequencies (0.2 THz) are predicted for systems
with contact radii on the order of $r_* = 3$ nm.  A surprising
prediction is the saturation and red shift of frequency for currents
large enough to invert the magnetization under the point contact.
This is a direct consequence of the exchange term included in our
model.

The magnetization $\vecM = (M_x,M_y,M_z)$ in the free layer can be
described by the vectorial equation {\setlength\arraycolsep{1pt}
\begin{eqnarray}
  \pd{\vecM}{\tau} & = &\underbrace{- \abs{\gamma} \mu_0 \vecM \times
    \vec{H}_{\text{eff}}}_{\text{precession}} -
  \underbrace{\frac{2}{\MS^2 T_2} \vecM \times (\vecM \times
    \vec{H}_{\text{eff}})}_{\text{Landau-Lifshitz damping}} +
  \nonumber\\[-2mm] 
  \label{eq:13}
  & & \underbrace{\beta(\vec{x}) \vecM \times (\vecM \times
    \hat{m}_{\text{F}})}_{\text{Slonczewski SMT torque}},
\end{eqnarray}}
$\!\!$where the precessional and damping terms are driven by the
effective magnetic field 
  $\vec{H}_{\text{eff}} = H_0 \hat{z} - M_z \hat{z} 
  + \frac{D}{\abs{\gamma} \mu_0 M_s \hbar} \nabla^2 \vecM$. 
This field consists of the applied magnetic field, $H_0 \hat{z}$, the
demagnetizing field due to axial dipole coupling, $-M_z \hat{z}$, and
the exchange field proportional to $\nabla^2 \vecM$
\cite{LandauLifshitzStatPhys2}.  The Landau-Lifshitz form of damping
is a commonly used phenomenological term that drives the magnetization
to align with the total field.  The Slonczewski SMT torque term is
derived in \cite{Slon96} and assumes a large, bulk ferromagnetic layer
with fixed magnetization direction $\hat{z}$.  Relevant parameters are
the gyromagnetic ratio ($\gamma\! =\! g \mu_B/\hbar$), the spectroscopic
splitting factor ($g$), the Bohr magneton ($\mu_B$), Planck's constant
divided by $2\pi$ ($\hbar$), the free space permeability ($\mu_0$),
the exchange coupling parameter ($D$), the magnetization volume
density at saturation ($M_s$), and the transverse relaxation time
($T_2$); $\beta(\vec{x})$ is a driving term depending on the current
discussed below.

We assume zero temperature and negligible crystalline anisotropy in
order to gain insight into the basic mode structure facilitated by the
fundamental physics associated with SMT in point contacts.  Such
effects can be accounted for by modification of eq \eqref{eq:13} to
include a random fluctuation term and an effective anisotropy field.
Also, since the thickness of the free layer is small, $\delta \approx
5$ nm, the in-plane components of the dipole field contribution to
$\vec{H}_{\text{eff}}$ can be neglected \cite{Schneider04}.

The magnetization is assumed to have rotational symmetry: the spatial
variation of the magnetization depends solely on the distance $r$ from
the center of the point contact.  Then, the current term, $\beta$, is
given by
\begin{equation*}
  \beta(r) = \frac{I \hbar \epsilon \gamma}{2 \MS^2 \delta \pi r_*^2 e}
  \Phi(r_*\! -\! r), ~ 
  \Phi(r_*\! -\! r) = \left \{
    \begin{array}{cc}
      1 & r \le r_* \\
      0 & r > r_*
    \end{array} \right.,
\end{equation*}
where $\epsilon$ is the SMT efficiency, $\delta$ is the thickness of
the free layer, $e$ is the charge of an electron, $r = (x^2\! +\!
y^2)^{1/2}$, and $\Phi(r_*\!  -\! r)$ is the Heaviside step function
defining the point contact to be a circle of radius $r_*$.  The
efficiency $\epsilon$ is in principle a complicated function of many
parameters, including (but not limited to) microscopic details of the
interfaces in question \cite{Slon99,Stiles02}.  Many of the
approximations used in the calculation of $\epsilon$ ignore the
lateral geometry of the actual experiments, choosing instead to use a
more tractable 1D calculation.  Since $\epsilon$ is difficult to
approximate, let alone calculate, and is not the subject of this
paper, we will treat $\epsilon$ as a constant fitting parameter.

By taking the dot product of equation \eqref{eq:13} with $\vecM$, one
sees that the magnetization of the free layer is locally conserved for
all times, i.e., $|\vecM| = \MS = $ constant, or
\begin{equation}
  \label{eq:25}
  \MS^2 = M_x^2(r,\tau) + M_y^2(r,\tau) + M_z^2(r,\tau),
  ~ r \ge 0, ~ \tau \ge 0. 
\end{equation}
Consider the standard normalization, $\vm\! =\! \vec{M}/\MS \! = \!
(m_x,m_y,m_z)$, $t\! =\! \omega_M \tau$, $\rho = r/\;l_{\text{ex}}$,
where $\omega_M \equiv \gamma \mu_0 M_s$ and the exchange length is
$l_{\text{ex}} \equiv \sqrt{D/\gamma \mu_0 M_s \hbar}$.  For the
analysis, it is convenient to encode the transverse components of
$\vm$, $\vm_{\perp} = (m_x,m_y)$, in the complex quantity $m = m_x + i
m_y$.  The axial component of the magnetization is treated as a
perturbation from the equilibrium solution $\vm \equiv \hat{z}$, so
$m_z = 1 - f$, where $f$ is the normalized perturbation.  Evaluating
the vector cross products in \eqref{eq:13} leads to
{\setlength\arraycolsep{1pt}
\begin{eqnarray}
  i \pd{m}{t} &=& m \nabla^2 f + (1-
  f+i\alpha)\nabla^2 m + (1-f-h)m \nonumber \\[-2mm]
  & &+ i\bigg\{j
  \Phi(\rho_* - \rho) - \alpha(h-(1-f))\bigg\}(1-f)m \nonumber \\[-2mm]
  & &+ i \alpha \bigg\{ \abs{\nabla f}^2 + \abs{\nabla m
  }^2
  \bigg\}m \nonumber \\[-2mm]
  \pd{f}{t} &=& \im (m^* \nabla^2 m)\! +\! \Big\{j
  \Phi(\rho_*\! 
  - \!\rho) \!-\! \alpha(h \!-\! (1\!-\!f)) \Big\}\abs{m}^2 \nonumber \\[-2mm]
  \label{eq:1}
  & &+ \alpha\Big\{ \nabla^2 f - (\abs{\nabla m}^2 +
  \abs{\nabla f}^2)(1-f) \Big\},
\end{eqnarray}}
$\!\!$where $\nabla^2 \equiv \pdd{}{\rho} + \frac{1}{\rho} \pd{}{\rho}$. 
The dimensionless parameters for applied field, damping, and current
are respectively
\begin{equation*}
  h = \frac{H_0}{M_s}, ~ \alpha = \frac{2}{\gamma \mu_0
    M_s T_2}, ~ j = \frac{\hbar \epsilon}{2 M_s^2 e \mu_0 \pi
    r_*^2 \delta}I. 
\end{equation*}
The coupled system (\ref{eq:1}) is equivalent to the model
\eqref{eq:13}.  Since the patterned multilayer mesa used in
experiments is two orders of magnitude wider than the point contact
\cite{Silva04}, we assume that the multilayer has infinite extent in
the $xy$ plane, hence $0 \le \rho < \infty$.

The \emph{nominal} parameter values used in this paper are $r_* \!=\!
20$ nm, $\epsilon \!=\! 0.26$, $D \!=\! 4
\,\text{meV}\!\cdot\!\text{nm}^2$, $M_s \!=\! 640$ kA/m, $\delta \!=\!
5$ nm, $\gamma \!=\! 1.85\! \cdot\! 10^{11}$ Hz/T, $\omega_M \!=\!
23.68$ GHz, $l_{\text{ex}} \!=\! 6.40$ nm, and $\alpha \!=\!  0.0112$.
We show that by using only $\epsilon$ as a fitting parameter, the
theoretical data match experiment well.  Except where noted, the
parameter values $h \!=\! 1.1$ and $\rho_* \!=\! 3.12$ are used.

In order to develop a perturbation scheme, we introduce a small
parameter, $\frac{a^2}{2}\! \ll\! 1$, and a rescaling of equation
(\ref{eq:1}), $m \!=\! a \tilde{m}$.  The parameter $a$ represents the
magnitude scale of the transverse magnetization at the center of the
point contact.  Using the approximation, $f \approx \frac{a^2}{2}
|\tilde{m}|^2$ for the saturation condition, substituting this
approximation into eq \eqref{eq:1}, and keeping only first order terms
in $a^2$, we find that $\tilde{m}$ satisfies a CGL-type equation,
{\setlength\arraycolsep{1pt}
\begin{eqnarray}
  i \pd{\tilde{m}}{t} &=& (1+i\alpha)\nabla^2 \tilde{m} -
    (h-1)\tilde{m} +i \Big\{j\Phi -
    \alpha(h-1) \Big\}\tilde{m} \nonumber \\[-2mm]
  & &+\frac{a^2}{2} \Bigg\{\!\!\underbrace{-\abs{\tilde{m}}^2 \tilde{m}
  }_{\text{dipole coupling}} \!\! + i\Big[j 
  \Phi - \alpha (h-2) \Big] \abs{\tilde{m}}^2 \tilde{m}
  \nonumber \\[-2mm]
  \label{eq:2}
  & &+ \underbrace{\Big[ \tilde{m} \nabla^2 \tilde{m}^* + 2
    (1+i\alpha) \abs{\nabla \tilde{m}}^2 \Big]
    \tilde{m}}_{\text{nonlinear exchange}} \Bigg\}. 
\end{eqnarray}}
\!\!\!Seeking a steady-state solution of the form $\tilde{m}(\rho,t) 
= e^{i\hat{\omega} t} \phi(\rho)$, $\hat{\omega} = h - 1 - \omega$
yields a \emph{nonlinear} eigenvalue problem with boundary conditions
$\phi(0) \!=\! 1$, $d\phi/d\rho(0) \!=\! 0$, $\lim_{\rho\to\infty}
\phi(\rho) \!=\! 0$.  The corresponding \emph{linear} eigenvalue
problem was solved by Slonczewski \cite{Slon99} with two real
eigenvalues $\omega$ and $j$.  This suggests the asymptotic expansions
\begin{equation}
  \label{eq:15}
  \begin{split}
    &\phi(\rho) = \phi^{(0)}(\rho) + \frac{a^2}{2} \phi^{(1)}(\rho) +
    \cdots \\ 
    \omega = \omega^{(0)} + &\frac{a^2}{2} \omega^{(1)} + \cdots,\quad
    j = j^{(0)} + \frac{a^2}{2} j^{(1)} + \cdots.
  \end{split}
\end{equation}
Slonczewski's linear result is summarized below.\\
\textbf{Linearized solution \cite{Slon99}:} \textit{The leading order
  mode $\phi^{(0)}(\rho)$ in the expansion (\ref{eq:15}) is}
\begin{equation*}
  \phi^{(0)}(\rho) = \left\{
    \begin{array}{lc}
      J_0(k_i \rho) & 0 \le \rho \le \rho_* \\
      c H_0^{(2)}(k_o \rho) & \rho_* < \rho
    \end{array}
  \right. ~,\quad 
  c = \frac{J_0(k_i \rho_*)}{H_0^{(2)}(k_o \rho_*)},
\end{equation*}
$k_i = (\frac{-\omega^{(0)} +i(j^{(0)}-\alpha(h-1))}
{1+i\alpha})^{1/2}$, $k_o =(\frac{-\omega^{(0)}-i\alpha(h-1)}
{1+i\alpha})^{1/2}$, \textit{where the linear eigenvalues $\omega^{(0)}$ and
  $j^{(0)}$ are determined by the real and imaginary parts of}
\begin{equation}
  \label{eq:17}
  k_i H_0^{(2)}(k_o \rho_*) J_1(k_i \rho_*) = k_o H_1^{(2)}(k_o
  \rho_*) J_0(k_i \rho_*). 
\end{equation}

Equation (\ref{eq:17}) is solved numerically using a nonlinear root
finder.  A doubly-discrete set of eigenvalues for frequency and
current is found.  We consider only the lowest eigenvalues
$\omega^{(0)}$ and $j^{(0)}$, where $j^{(0)}$ is the \emph{critical
  current} for linear spin wave excitations \cite{Slon99}.

The predicted frequency of the ground state mode, $\hat{\omega} \sim h
- 1 - \omega^{(0)}$, is valid only for very small amplitudes $a$,
whereas experiments produce much higher frequencies \cite{Silva04}.
The nonlinearity in \eqref{eq:2} gives rise to a frequency shift
which, through the use of the Poincar\'{e} method \cite{Cole68}, helps
explain this disparity.
\\
\textbf{Nonlinear solution:} \textit{The first order nonlinear
  frequency and critical current shifts $\omega^{(1)}$ and $j^{(1)}$
  are determined by the linear system}
\begin{equation*}
  \left[
    \begin{array}{cc}
      \re(I_2) & \im(I_1) \\
      \im(I_2) & -\re(I_1)
    \end{array}
  \right]
  \left[
    \begin{array}{c}
      \omega^{(1)} \\
      j^{(1)}
    \end{array}
  \right] =
  \left[
    \begin{array}{c}
      \re(I_3) + \im(I_4) \\
      \im(I_3) - \re(I_4)
    \end{array}
  \right],
\end{equation*}
\textit{where $I_k$ depend only on the leading-order solution:}
\begin{eqnarray*}
  I_1 &\equiv& \frac{\rho_*^2}{2} \bigg\{ J_0^2(k_i \rho_*) +
  J_1^2(k_i \rho_*) \bigg\} \\[-2mm]
  I_2 &\equiv& I_1 -\frac{c^2 \rho_*^2}{2} \bigg\{ (H_0^{(2)}(k_o
  \rho_*))^2 + (H_1^{(2)}(k_o \rho_*))^2 \bigg\} \\
  I_3 &\equiv& \int_0^\infty \!\bigg\{\!\!-|\phi^{(0)}|^2\! +\! \phi^{(0)}
  \nabla^2 \phi^{(0)*}\! +\! 2 |\nabla \phi^{(0)}|^2 \bigg\}
  \phi^{(0)^2} \rho \,\ud\rho \\
  I_4 &\equiv& \int_{0}^\infty \!\bigg\{ \alpha(2-h) |\phi^{(0)}|^2 - 2
  \alpha |\nabla \phi^{(0)}|^2 \bigg\} \phi^{(0)^2} \rho \,\ud\rho \\[-2mm]
  & & + \,j^{(0)} \int_0^{\rho_*} |\phi^{(0)}|^2 \phi^{(0)^2} \rho \,\ud\rho. 
\end{eqnarray*}

The above perturbation method provides a means for calculating the
nonlinear frequency and current shifts, $\omega^{(1)}$ and $j^{(1)}$,
of a mode solution to \eqref{eq:2} and helps explain experimental
results \cite{Tsoi98}.  The frequency and critical current have a
parabolic dependence on the transverse amplitude $a$
\begin{equation}
  \label{eq:37}
  \hat{\omega} \sim h - 1 -\omega^{(0)} - \frac{a^2}{2}
  \omega^{(1)}, \qquad j \sim j^{(0)} + \frac{a^2}{2} j^{(1)}. 
\end{equation}
We note that using equations (\ref{eq:37}), the relation between
$\hat{\omega}$ and $j$ is linear because both scale as $a^2$. 

The first-order nonlinear contributions from equation \eqref{eq:2} due
to exchange and dipole coupling were compared.  Using equations
\eqref{eq:37}, the frequency and current dependence as parameterized
by $a$ were calculated when the exchange and dipole coupling
nonlinearities were removed from the calculation.  We find that if
nonlinear exchange terms are ignored, the relative error in the slope
$d\hat{\omega}/dj$ is approximately $10\,\%$ for $r_* = 20$ nm,
whereas for $r_* < 12$ nm the relative error is greater than $25\,\%$.

Experiments \cite{Silva04} show that for large currents, and hence
larger amplitudes, the frequency/current relationship significantly
deviates from linearity.  In order to account for this, equation
(\ref{eq:1}) was solved numerically using the projection method where
one renormalizes the magnetization after each time step in order to
preserve the local conservation of $\vecM$ (\ref{eq:25}) (see
\cite{Weinan00}).  The steady states of this parametrically
forced/damped system are attractors.  The initial condition $m(\rho,0)
= a\phi^{(0)}(\rho)$ was evolved in time over a large domain
($0\!\le\!  \rho\! \le\! 100$) until the solution relaxed to a steady
state.  The mode frequency was numerically recovered from the phase of
the precessing transverse component as $\hat{\omega} = \frac{d}{dt}
\arg[m(0,t)]$, $t \to \infty$.

\begin{figure}
  \centerline
  {\includegraphics[scale=.35,angle=270]{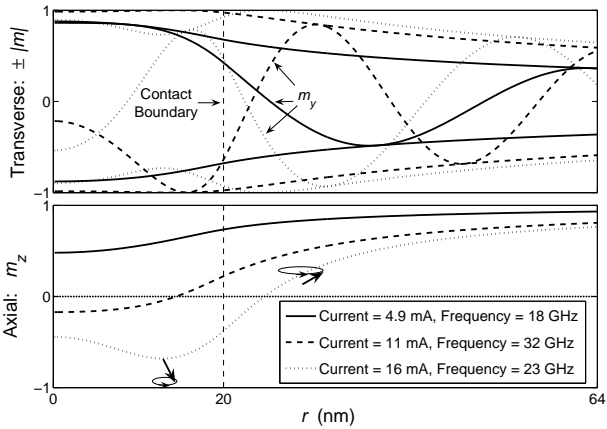}}
  \caption{Spatial dependence of fully nonlinear modes with excitation
    current and frequency.  Top: transverse envelopes $\pm|m| =
    \pm\sqrt{m_x^2+m_y^2}$ and associated spin waves $m_y$.  Bottom:
    axial components $m_z$.  For $j \ge j_{max}$, the transverse mode
    magnitude is no longer monotonically decreasing since the axial
    component of magnetization is negative ($m_z < 0$) in the point
    contact.  Frequency and current values for each mode are shown as
    circles in fig.  \ref{fig:freq_current_expt}.}
  \label{fig:nl_modes}
\end{figure}
Typical modes, their frequencies, and currents are shown in fig.
\ref{fig:nl_modes}.  The transverse mode envelope $\pm \abs{m} =
\pm(m_x^2+m_y^2)^{1/2}$ and the $y$-component of magnetization $m_y$
show slow decay, which means that spin waves are continually radiating
away from the point contact, and hence energy from the contact region
is lost into the surrounding medium.

\begin{figure}[h]
  \centering 
  \includegraphics[scale=.4,clip=true,trim=0 5 0 0]{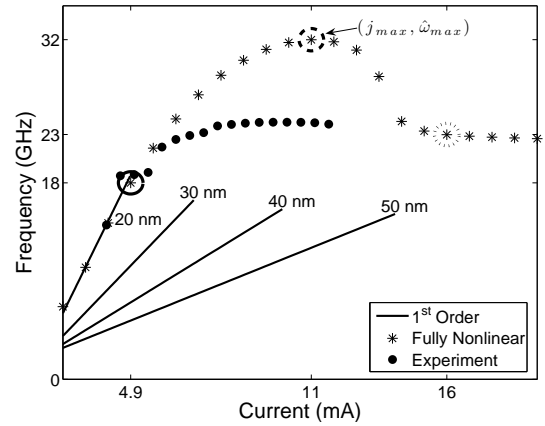}
  \caption{Precession frequency as a function of current with
    comparison between fully nonlinear modes (direct numerical
    simulation of eq (\ref{eq:1})), the perturbation result
    \eqref{eq:37}, and experiments conducted using $r_*=20$ nm point
    contacts \cite{Silva04}.  Large circles represent the frequency
    and current of modes plotted in fig. \ref{fig:nl_modes}.  The
    straight lines were generated by perturbation theory (eq
    \eqref{eq:37}) for the contact radii denoted.  Note the excellent
    agreement between numerical simulation, perturbation theory, and
    experiment for small currents.  The parameters have nominal values
    except $\epsilon=0.8$.}
  \label{fig:freq_current_expt}
\end{figure}

The plot of frequency versus current in fig.
\ref{fig:freq_current_expt} shows a comparison between numerical
simulation of \eqref{eq:1}, perturbation theory \eqref{eq:37}, and
experiment \cite{Silva04}.  For current excitations larger than
$j_{max}$, the solitary-wave mode structure becomes highly nonlinear
(compare with fig.  \ref{fig:nl_modes}).  There is a maximum frequency
and corresponding current for a given physical system
$(j_{max},\hat{\omega}_{max})$.  As the dashed mode in fig.
\ref{fig:nl_modes} shows, frequency saturation corresponds to
magnetization precession along the equator near the center of the
point contact ($m_z \approx 0$).  For currents less than $j_{max}$,
the transverse mode magnitude has a monotonically decreasing
dependence on $r$.  For larger currents, the mode is not monotonic and
develops interesting structure inside the point contact.

Fig. \ref{fig:freq_current_expt} shows that, as the point contact size
is increased, the slope of frequency versus current decreases because
the slope depends on SMT torque (inversely proportional to $r_*^2$)
and damping due to spin wave generation (inversely proportional to
$r_*$ in the linear approximation \cite{Slon99}).  The comparison of
the simulated results (\textasteriskcentered) with experiment is
remarkable when the SMT efficiency is taken to be $\epsilon\! =\!
0.8$.  As the experimental values of the frequency level out, full
saturation is being reached.  This is the first model to qualitatively
capture this type of behavior.

Nominal values for $\epsilon$ are on the order of $0.25$ for most
ferromagnetic metals \cite{Pufall03}.  The fact that $\epsilon > 0.25$
is used to fit the experimental data suggests that further refinement
of the model is desirable.

\begin{figure}
  \centering
  {\includegraphics[scale=.44,clip=true,trim=0 0 0 5]{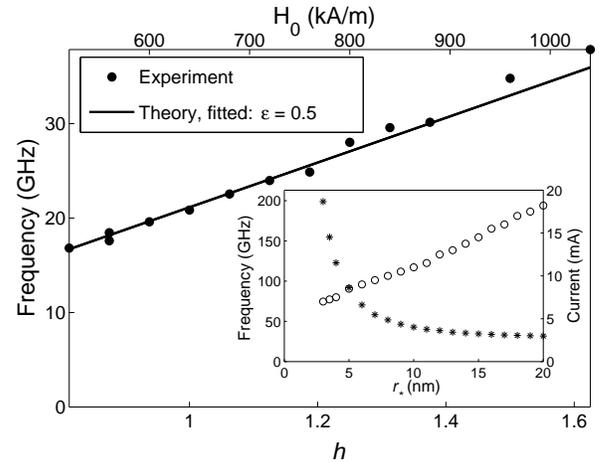}}
  \caption{Frequency as a function of applied field $h$.  Comparison
    between fully nonlinear modes (direct numerical simulation of eq
    (\ref{eq:1})) and experiments \cite{Silva04}.  Parameters are
    $I=10$ mA and fitted $\epsilon=0.5$.  Inset: maximum frequency
    (\textasteriskcentered) and current ($\circ$) versus contact
    radius.}
  \label{fig:f_vs_H}
\end{figure}
The dependence of frequency on applied magnetic field $H_0$ is
depicted in fig. \ref{fig:f_vs_H}.  With the fitted SMT efficiency
parameter, the theory compares well with experiment.  The mode
structure is highly nonlinear because the transverse amplitude is
$|m(0,t)| \approx 0.98$, which precludes the use of eq \eqref{eq:37}.
However, a full numerical solution is required for only one particular
value of $h$.  Since $d\hat{\omega}/dh \approx$ constant for $\alpha\!
\ll\! 1$, once the parameters $\alpha$, $j$, and $\rho_*$ are chosen,
the y-intercept for the approximate linear relationship between
$\hat{\omega}$ and $h$ is fixed.  The slope of the line does not
change.  The theory predicts the slope very accurately.  In fig.
\ref{fig:f_vs_H} inset, much higher frequencies are predicted by
shrinking the size of the point contact.  Using the method stated
above, we predict that frequencies in the 0.2 THz range are attainable
for a point contact with a radius of 3 nm.  We find that as the point
contact is made smaller, the current decreases linearly for the same
reasons stated earlier in regard to the $\hat{\omega}(j)$ slope, while
the frequency increases in proportion to $1/r_*^2$ due to
exchange-mode spin-wave dispersion.

In summary, a new model of magnetic excitations in point contact
structures that includes nonlinearities due to both exchange and
dipole coupling is introduced.  Spatially dependent steady-state modes
are calculated.  This represents a large step forward in the
explanation of the phenomena observed in recent experiments.

\begin{acknowledgments}
  This work was supported by NSF grants DMS-0303756, VIGRE
  DMS-9810751, AFOSR grant F-4620-03-0250, and the DARPA SPinS
  program.
\end{acknowledgments}


\bibliographystyle{apsrev}

\end{document}